\documentclass[conference]{IEEEtran}
\usepackage{balance}
\usepackage{graphicx}
\usepackage{booktabs}
\usepackage{multirow}
\usepackage[abbreviations]{foreign}
\usepackage[inline]{enumitem}
\PassOptionsToPackage{hyphens}{url}
\usepackage{hyperref}
\usepackage{pifont}
\usepackage{subcaption}

\hypersetup{
  colorlinks = true, 
  urlcolor   = blue, 
  linkcolor  = blue, 
  citecolor  = red   
}

\newcommand{\sys}{\textsc{PCRAFT}\xspace}

\begin{document}

\title{PCRAFT: Capacity Planning for Dependable Stateless Services}

\author{\IEEEauthorblockN{Rasha Faqeh\IEEEauthorrefmark{3}, André Martin\IEEEauthorrefmark{3}, Valerio Schiavoni\IEEEauthorrefmark{1}, Pramod Bhatotia\IEEEauthorrefmark{2}, Pascal Felber\IEEEauthorrefmark{1} and Christof Fetzer\IEEEauthorrefmark{3}}
 \IEEEauthorblockA{\IEEEauthorrefmark{1}University of Neuchâtel, Switzerland. E-mail: \url{first.last@unine.ch}}
 \IEEEauthorblockA{\IEEEauthorrefmark{2}Technical University of Munich, Germany. E-mail: \url{first.last@in.tum.de}}
 \IEEEauthorblockA{\IEEEauthorrefmark{3}Technical University of Dresden, Germany. E-mail: \url{first.last@tu-dresden.de}}
}


\maketitle

\sloppy

\begin{abstract}%
Fault-tolerance techniques depend on replication to enhance availability, albeit at the cost of increased infrastructure costs.  This results in a fundamental trade-off: Fault-tolerant services must satisfy given availability and performance constraints while minimising the number of replicated resources. These constraints pose {\em capacity planning} challenges for the service operators to {\em minimise replication costs without negatively impacting availability}.

To this end, we present \sys,\footnote{\textbf{P}erformant, \textbf{C}heap, \textbf{R}eliable and \textbf{A}vailable \textbf{F}ault \textbf{T}olerance.} a system to enable capacity planning of dependable services. \sys's capacity planning is based on a hybrid approach  that combines empirical performance measurements with probabilistic modelling of availability based on fault injection. In particular,  we integrate traditional service-level availability mechanisms (active route anywhere and passive failover) and deployment schemes (cloud and on-premises) to quantify the number of nodes needed to satisfy the given availability and performance constraints.Our evaluation based on real-world applications shows that cloud deployment requires fewer nodes than on-premises deployments. Additionally, when considering on-premises deployments, we show how passive failover requires fewer nodes than active route anywhere. Furthermore, our evaluation quantify the quality enhancement given by additional integrity mechanisms and how this affects the number of nodes needed.
\end{abstract}


\section{Introduction} 
\label{sec:introduction}

Dependability is a must-have requirement for modern Internet-based services.
These services must be \emph{highly available, integrity protected, secure} and offer high assurance of \emph{performance} to users, in terms of response time and throughput.
Service providers can use replication to achieve high availability and more reliable services, and deploy them either using an on-premises cluster or on a public cloud.
However, such techniques regardless of the chosen deployment
\begin{enumerate*}[label=\emph{(\roman*)}]
    \item increase the complexity and the costs of the supporting infrastructure, and
    \item degrade the observed performance of the applications.
\end{enumerate*}
To address the problem of performance assurance, the service providers provision extra physical resources.
However, over-provisioning increases the operational costs without improving availability.
This represents an important problem:
\emph{how to achieve the required level of availability and performance while maintain affordable costs?}
We tackle this problem directly by providing a capacity planning process.

We consider a scenario of a service that uses a cluster of server nodes that run stateless \cite{rodriguez2008restful} or soft-state \cite{birman2012overcoming} applications.
The service has a target throughput and availability requirements.
To fulfil these requirements, applications are replicated on different nodes.
However, nodes crash and are replaced with new ones, mainly to enhance or guarantee availability of the services.
This can be implemented using simple replication techniques such as \emph{active route anywhere} (ARA) or \emph{passive failover} (PF).
\emph{Active route anywhere} provisions more nodes to process users requests than required to meet the target performance.
The extra nodes fulfil this objective as long as sufficiently many stay available.
\emph{Passive failover} also requires extra nodes to be provisioned but they are passively waiting in a standby pool: when an active node fails, it is replaced by one from the pool.
After repairing the node, it is returned to the pool to handle new failures.

We consider two physical deployment schemes: \emph{on-premises cluster} and \emph{public cloud}. 
The scheme has direct consequences on the fault-tolerance properties as well as the associated costs.
Cloud providers largely rely on the passive failover approach, restarting virtual machines (or lately containers) of a failed node on a functional one.
Doing so, they can meet service level agreements (SLAs) under several classes of \emph{nines} \cite{greene2007carrier}. 
Note that the service provider only has to pay
for the nodes actively participating in the cluster without the need to pay for the nodes that passively exist in the pool.
Conversely, when compared to an on-premises deployment, resource sharing in the cloud (\eg co-located virtual machines) impacts the observed performance, leading to a degraded response time and unexpected variations \cite{runtimeVariationInTheCloud}.
This degradation is perceived by clients as service unavailability if it exceeds a certain (negative) threshold.
On the other hand, on-premises solutions must provide and pay for both the active nodes in the cluster and the passive nodes in the pool.

Even when the service is available and serves requests, because of transient hardware faults these requests may be incorrectly processed and produce incorrect results.
Therefore, mechanisms such as instruction level redundancy \cite{Reis2005} are often used to protect the integrity of the execution by detecting which executions are incorrectly processed, yet at the price of decreased performance and corresponding capacity.

Our system, \sys, proposes a capacity planning process that is based on a combination of empirical performance measurements with availability and integrity probabilistic modelling.
We consider the costs of deploying dependable applications in terms of both the number of nodes used and the integrity of the service provided.
This simpler cost model allows us to easily reason about the total cost of ownership of a cluster and to consider the number of nodes as a building block for more complex schemes.

\sys integrates two availability techniques in its process: \emph{passive failover} and \emph{active route anywhere}.
PF consists of loosely coupled and independent servers with failover capabilities.
To tolerate $f$ failures, at least $f$ extra nodes should be available in a standby pool.
By default, these nodes are in \emph{cold} mode, \ie they are only started when needed.
ARA uses active replication to deploy $f$ additional, fully functional nodes behind a load-balancer that dispatches requests to all of them.
The service performance is ensured only if at most $f$ nodes crash.
In addition, \sys uses \emph{instruction level redundancy} (ILR) \cite{Reis2005} to protect node integrity from transient faults.
ILR replicates data flow instructions and executes two instruction streams in parallel, leveraging the instruction-level parallelism of modern CPUs.
To do so, \sys relies on the HAFT \cite{Kuvaiskii2016HAFT} framework, which additionally exploits hardware transactional memory (HTM) \cite{IntelTSX2013} to recover from faults.

Previous studies on capacity planning mainly modelled the system performance with varying workloads and resource conditions \cite{CPWebApplications,DSML,CombinedImpact,perfromanceAssurance} assuming an always available reliable infrastructure.
Studies that quantify the availability in the event of different types of failures ignored the effect of failures on the performance \cite{DifferentiatedAvailability,availabilityVmNonVm}.
Similarly, the combined effect of faults on availability and performance \cite{BothEndToEnd,QOSAvailabilityPerformance,FaultInjectionAndModelingPerformability,PerformabilityOfCluster} did not quantify the number of nodes needed to ensure both the performance and availability levels.

In this paper, we propose the following contributions.
\begin{enumerate*}[label=\emph{(\roman*)}]
    \item We introduce \sys, a capacity planning process that is based on a combination of empirical experimentation and modelling to quantify the number of nodes needed to assure availability and performance levels.
    \item We develop a collection of probabilistic models to measure the availability of services when incorporated with various failures and recovery behaviour via different fault-tolerance approaches and physical deployment schemes.
    \item We measure the integrity of the service by combining integrity models with fault-injection.
\end{enumerate*}

The remainder of the paper is structured as follows.
\S\ref{sec:related} provides additional information on dependability schemes and related work on capacity planning processes.
We describe our fault model and present our approach in \S\ref{sec:design}.
Probabilistic models are discussed in \S\ref{sec:implementation}.
We show in \S\ref{sec:evaluation} how one can use \sys's capacity planning process to generate the associated costs, before concluding in \S\ref{sec:conclusion}.


\section{Related work} 
\label{sec:related}

We review dependability approaches as well as studies related to capacity planning.

\subsection{Dependability approaches}
Service dependability encompasses various measures.
We focus on integrity and availability.
Crash failures degrade the service availability.
Integrity failures degrade both the service reliability and availability.

\smallskip\noindent\textbf{Availability.}
Availability enhancing approaches attempt to harden against crash failures.
Crash failures implement fail-stop behaviour, where the faulty hardware component stops executing the algorithm.
A standard approach to achieve high availability is the use of state machine replication \cite{SMR1990}.
The server application is replicated actively on $\mathit{2f+1}$ nodes to tolerate $\mathit{f}$ nodes crashes.
If a minority of replicas crashes, the surviving replicas keep serving requests.
Due to the high runtime and infrastructure costs, this approach is usually restricted to protect stateful applications. 

In the context of stateless applications, highly available systems \cite{HAProxy} replicate server applications either using passive failover or active replication schemes. 
\emph{Passive failover} consists of loosely coupled and independent servers with failover capabilities which are connected to a monitoring service (\eg heartbeats).
To tolerate $f$ failures, at least $f$ extra nodes should be available in a standby pool.
The monitoring service detects server crashes and initiates a failover to restart/migrate the service using a node from the pool.
The failover time, \ie the time needed to replace the failed node and to reconfigure the new node, directly affects the availability of the service.
The failover can be implemented using a fast switchover to a \emph{hot} node from the pool, already started with the application configured and its state periodically updated (\eg VMware fault tolerance).
Alternatively, a delayed switchover to a \emph{cold} node not yet started is also possible (\eg VMware high availability).

In contrast, \emph{active replication} schemes deploy for the service $f$ additional, fully functional nodes behind a load-balancer that dispatches requests to all of them.
The service performance is ensured only if at most $f$ nodes crash.
The load balancer distributes requests in two ways.
With \emph{sticky} sessions, user requests are routed to a predetermined physical node for processing (user session data are cached locally), but possibly leading to load imbalances.
\emph{Active route anywhere} does not restrict requests to specific nodes, and load is equally balanced. 
However, user session data is not cached: locality information is lost, as the probability for subsequent requests to route to the same physical node is low.
\sys integrates both PF in cold mode and ARA in its process.

\smallskip\noindent\textbf{Transient faults.}
Hardware failures are not restricted to crashes.
Transient faults could manifest as arbitrary state corruptions \cite{PaxosMadeLive2007}, data loss \cite{AmazonLoadBalancer2008} and, possibly, outages \cite{AmazonS32008}.
Surprisingly, transient faults occur at high rates and reappear frequently after the first occurrence \cite{CyclesCells2011,hogpen}. 
Servers can protect the memory and caches using error correcting codes (ECC).
However, protecting the CPU computation and registers is hard.
Byzantine fault tolerance (BFT) \cite{pbft} protects against silent data corruptions, transient faults and malicious attacks.
However, BFT is not broadly adopted since it introduces considerable costs due to performance and management overheads \cite{Song09bftfor}.

Other hardening approaches attempt to reduce the replication and runtime overheads.
These approaches add redundancy locally at the level of processes, threads, and instructions.
Process level redundancy creates one slave redundant process for each application master process in the program, and synchronises state on system calls \cite{PLR2007}.
Redundant multi-threading \cite{DAFT} executes code in redundant threads instead of processes.
However, both approaches require deterministic multi-threading.
\emph{Instruction level redundancy} (ILR) \cite{Reis2005} replicates data flow instructions to create a shadow stream that shares the same memory state with the master stream but uses an exclusive copy of CPU registers.
The two streams execute in parallel, leveraging the instruction-level parallelism of modern CPUs.
The master stream updates the memory after comparing the memory update results with the shadows.
ILR does not restrict threads interleaving and hence allows for non-determinism in applications.
However, ILR replicates at instruction level: fault detection is limited to faults in the CPU registers or execution unit.
HAFT \cite{Kuvaiskii2016HAFT} is a state-of-art framework that hardens server applications against data corruptions using instruction level redundancy \cite{Reis2005} to detect faults and hardware transactional memory (HTM) \cite{IntelTSX2013} to recover from them.
It extends the LLVM compiler framework \cite{Lattner2004} to create self-checking applications by replicating the application data instructions and inserts periodic integrity checks wrapped inside HTM-based transactions.
Any mismatch between the two execution streams is revealed by the inserted integrity checks and the execution of the application is then flagged as corrupted.
\sys uses the HAFT approach to protect node integrity.

\subsection{Dependability-costs studies}
Capacity planning requires estimating the adequate size of different resources to ensure the dependability of the service.
Modelling techniques are well suited to address this problem, as we survey next.

\smallskip\noindent\textbf{Performance.}
Several studies use queuing theories and networks to predict the workload demands on resources such as CPU, memory and network bandwidth to support future computational needs \cite{DSML,perfromanceAssurance}.
Similarly, \cite{CPWebApplications,CombinedImpact} focus on workload characterisation of the arrival and service rates to estimate the throughput and response time. 
These studies use white-box analytical models to represent the server hardware/software and take into account the different resources of the system (CPU, memory, disk, threads, etc). 
As analytical models require domain expert knowledge and a large number of factors make the models complex, only those factors with the highest impact on performance, \ie which can tolerate minor errors, are considered in the performance predication.

Machine-learning based approaches \cite{BlackBoxForPerformanceCapacity,deployApplicationAndMeasurePerformance} adapt a black-box approach (\emph{i.e.}, zero-knowledge about the system) using empirical models.
While these models provide more accurate performance predictions, they are not portable across different hardware and deployments.
Similarly, \sys uses empirical measurements to predicate the service performance using specific software and hardware node configurations.

\smallskip\noindent\textbf{Availability.}
Several studies focused on estimating the effect of failures using stochastic models.
Failures happen at different granularities:
\begin{enumerate*}[label=\emph{(\roman*)}]
    \item CPU, memory \cite{availabilityVmNonVm},
    \item operating system, power supply, hypervisor and VM \cite{COAResourceEstimation,AvailabilitySRN}, or
    \item cluster, data-centre and cloud provider \cite{AvailabilityModelInDSN,DifferentiatedAvailability}.
\end{enumerate*}
State and non-state space modelling are two main approaches.
State space models, such as continuous-time Markov models (CTMC), stochastic Petri nets and stochastic networks \cite{DifferentiatedAvailability,availabilityVmNonVm,COAResourceEstimation,AvailabilitySRN} depict the system failure and repair behaviour using the states and rates to represent occurrence of events.
Non-state models \cite{COAResourceEstimation,AvailabilityModelInDSN}
capture the conditions leading to failures considering the structure of components.
Computationally-wise, these are usually cheaper.
In \sys, we focus on service unavailability regardless of which component failure caused the node to crash.
We use CTMC to represent \sys's availability models and capture the dependency between components, such as automated repair for components and failover in the cloud deployment.

\smallskip\noindent\textbf{Performability.} 
Performability studies model the effect of (un)availability and performance on a single metric.
They can take the form of a monolithic model, or a hierarchical model to reduce the
possibly generated state space.
\sys uses a hybrid approach for performability, which allows us to construct availability models that can be fed with  performance results from the empirical measurements instead of using performance modelling.
In \cite{BothEndToEnd}, authors quantify the effect of workload changes (\eg arrival and job service rate),  fault-load (physical machine failure rates) and system capacity on service quality when deployed in the cloud.
They consider the unavailability and the service response time as two different quality-of-service (QoS) metrics.
They use CTMC models and include both warm \cite{VOD} or cold \cite{DifferentiatedAvailability} standby nodes.
To quantify the QoS metrics, they assume the same number of nodes in each pool (warm/cold), scaling both pools by the same factor.
\sys can scale each pool independently to match certain performance and availability criteria.
As opposed to our proposal, they do not support active replication.
In \cite{QOSAvailabilityPerformance}, authors propose a hierarchical analytical model for both availability and latency, including correlated failures.
Their model can be leveraged to quantify the service response time and the success probability of client requests without using queueing models to predict the latency.
Additionally, they quantify the effect of adding more VMs to the service on the response time and service unavailability, with an increasing arrival request rate.
\sys tackles this problem by adding more nodes to cope with both performance and availability requirements. 


\section{System design} 
\label{sec:design}

We discuss in this section our assumptions about the fault model and we present the overall architecture and operation mode of \sys.

\subsection{Fault model}
We assume a dependable service built using a cluster of server nodes with three sources of failures:
\begin{enumerate*}[label=(\arabic*)]
    \item Availability failures at the node level regardless of the cause.
    We assume fail-stop failures that either crash nodes or make them non-responsive, \eg failing hardware, software, disks, memory.
    We also assume that server nodes fail independently but do not exclude that many nodes can fail simultaneously.
    We exclude failure types that cause multiple nodes to fail due to a single failure, \eg network failures or software bugs that deterministically crashes applications.
    \item Performance failures: requests cannot be served due to limited capacity, \ie the number of available nodes is too low.
    \item Integrity failures: nodes are available but requests are served incorrectly without any error notification, in particular silent data corruption (SDC) \cite{Reis2005} which leads to non-malicious Byzantine behaviour.
    SDCs are transient hardware faults caused by single event upsets, \eg one bit-flip in a CPU register or miscomputation in a CPU execution unit.
    Note that memory and caches are protected against bit-flips due to the use of ECC which is assumed to exist in all modern servers.
\end{enumerate*}

A service is \emph{dependable} only if
\begin{enumerate*}[label=\emph{(\roman*)}]
    \item the availability is above a predetermined threshold,
    \item whenever the service is available, there is sufficient capacity to ensure performance, and 
    \item the service data integrity is protected.
\end{enumerate*}

\subsection{Architecture}
We consider a distributed $n$-tier service, with a front-end load balancer, a middle-tier cluster of server nodes and a connected back-end persistent storage. 
We focus on the middle-tier (\ie web, application, caching servers) to be available and well-performing. 
We assume the load balancer to be always available (\eg using replication \cite{HAProxy}) and the back-end resources to be able to scale accordingly.

The load balancer intercepts the client requests and distributes them to the homogeneous cluster of servers nodes.
A node becomes saturated when serving the maximum throughput (requests per unit time) with an acceptable latency (application-dependent response time threshold). 
Adding more requests to the node after saturation will cause the latency perceived by users to be monotonically increasing as more requests are added to the node's local queue (sharing model \cite{QOSAvailabilityPerformance}). 
Alternatively, the server prematurely rejects the requests if the server is saturated---assuming no queues (constant bit rate model \cite{QOSAvailabilityPerformance})---with users subsequently re-sending the rejected requests.
In \sys, we adapt the constant bit rate model for server nodes to control the upper bound of the latency perceived by users.

\subsection{Dependable services}
The service consists of a cluster of nodes.
Its throughput is the sum of the throughput of each node, while the latency is the average latency of all nodes that constitute the cluster.
The service is available if at least a single node in the cluster is available, but with a degraded performance.
To consider the service dependable, it must fulfils both the availability constraints (\eg three 9's) and the performance constraints (\eg target throughput and response time threshold).
Additionally, the integrity mechanisms must be implemented at the node level to enhance the integrity of the served requests.

To meet our performance requirements, we can predict---based on single node performance---the number of nodes needed. Typically, we first assume that all nodes are always available. Since nodes can actually fail, in a second step we increase the availability by over-provisioning the number of nodes in the cluster. 
Over-provisioning might take the form of ARA in which fully functional extra nodes are added to the service cluster, or simply by replacing a failed node from a standby pool of nodes with PF. 
The number of nodes needed for the over-provisioning varies based on the type of deployment---on-premises or in the cloud.
This is due to the availability mechanisms which are implemented in the cloud to assure the availability level for a single node that must meet the SLA.
However, single node provisioning is typically insufficient to meet the availability constraints of a dependable service.
Finding the adequate number of nodes in each case requires solving the capacity planning problem for this service.

\subsection{Capacity planning process}
The process consists of two-phases.
First, we empirically benchmark the performance of a single server node.
Second, we model the availability behaviour and parameterise the models using the values from the first phase to identify how many server nodes we need to ensure the dependability requirements.

\smallskip\noindent\textbf{Experimentation phase.}
Assuming an always available node, we profile the service that consists of a single node along two dimensions: performance and integrity.
First, we benchmark the performance of both the native application ($\mathit{native}$) and the integrity protected version of the application ($\mathit{ft}$).
We refer to the throughput of the single node in both cases as $\mathit{NdT}_ \mathit{native}$ and $\mathit{NodT}_\mathit{ft}$, respectively.
Note that usually $\mathit{NodT}_\mathit{ft} < \mathit{NodT}_ \mathit{native}$.
The service target throughput $\mathit{SerT}$ is defined by the service provider as ($\mathit{Num}_\mathit{native} \times \mathit{NodT}_ \mathit{native}$) or ($\mathit{Num}_\mathit{ft} \times \mathit{NodT}_ \mathit{ft}$). 
For different node types ($\mathit{native}$ and $\mathit{ft}$), we can calculate $\mathit{Num}_\mathit{native}$ and $\mathit{Num}_ \mathit{ft}$ as an estimation for the number of nodes needed in the cluster to fulfil the service target performance assuming nodes do not fail and without any over-provisioning of nodes.
Note that usually $\mathit{Num}_\mathit{ft} > \mathit{Num}_\mathit{native}$ to fulfil the same  $\mathit{SerT}$.
Second, we inject integrity faults with a service that uses $\mathit{native}$ and $\mathit{ft}$ nodes to benchmark the integrity behaviour in each case. 
The results of the fault injection is subsequently used to estimate the integrity behaviour of applications in the presence of integrity faults.
                                                                                   
\smallskip\noindent\textbf{Modelling phase.}
We build availability models for a cluster of nodes that implements over-provisioning using ARA or PF mechanisms with a cloud or on-premises deployment.
The model can be used to first estimate the availability level achieved by the basic number of nodes $\mathit{Num}$ calculated in the previous phase. 
Then, by comparing the availability achieved to the target availability defined by the service provider, we  over-provision iteratively the number of nodes in the cluster until the availability constraint is met.
In addition to availability models, we build integrity models for $\mathit{native}$ and $\mathit{ft}$ to estimate the integrity behaviour of nodes in the cluster when deployed in the cloud.


\section{Implementation} 
\label{sec:implementation}

In the second phase of the process, we build continuous-time Markov chain (CTMC) models a probabilistic model checker tool called PRISM \cite{PRISM}.
The models, which capture the availability and the integrity behaviour, consist of possible states and possible transitions between them with their respective rates (failure and repair rates).
We assume all rates have inter-event times that are exponentially distributed \cite{exponentialDisGood}.
CTMC suffers from well-known state space explosion problems.
To reduce the generated number of states, we follow two approaches. 
First, we attempt to aggregate states together. Specifically, we use homogeneous nodes in the cluster and so they exhibit similar failure and repair behaviour.
It is thus not necessary to distinguish between which node in the cluster failed.
Rather, we can simply keep track of up nodes ($\mathit{UpNodes}$).
Second, performability analysis generally uses a hierarchy of models instead of monolithic ones in order to combine both availability and performance and hence reduce the number of generated states.
In this paper, instead of solving performance models, we use the empirical measurements to feed the availability models, which further reduces the number of states needed.

For our models, we explore the following server node types:
\begin{enumerate*}[label=\emph{(\roman*)}]
    \item $\mathit{native}$ nodes without any software/hardware hardening mechanisms implemented at the node level, and
    \item $\mathit{ft}$ nodes which are integrity protected using the HAFT \cite{Kuvaiskii2016HAFT} approach.
\end{enumerate*}
Additionally, we consider two deployments, in the \emph{cloud} and \emph{on-premise}, as well as two over-provisioning techniques, \emph{active route anywhere} and \emph{passive failover}.

\begin{figure}[t!]
	\centering
	\includegraphics[scale=0.7]{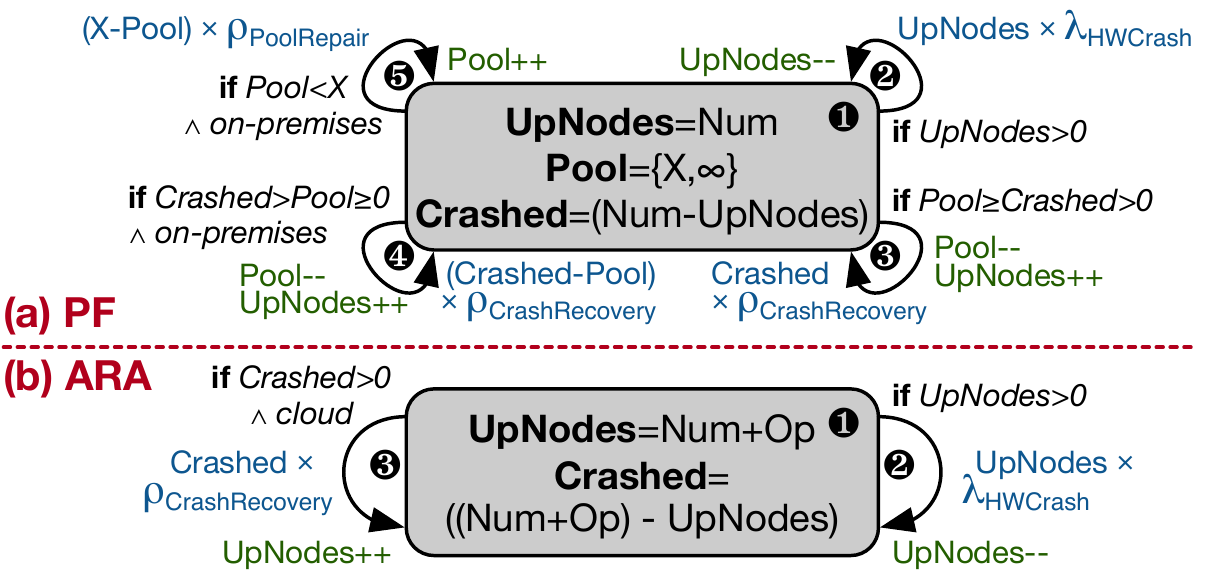}
	\caption{State machine for a cluster of nodes using passive failover (top) or active route anywhere (bottom) as over-provisioning in the availability models for the cloud and on-premises deployments.
    Rectangles represent the states while arrows are transitions with associated rates.
    The system fails at rates of $\lambda$'s and recovers at rates of $\rho$'s.}
    \label{fig:prism_availability_stateMachine}
\end{figure}

\smallskip\noindent\textbf{Availability models.}
In the availability models (\autoref{fig:prism_availability_stateMachine}), we inject hardware crash faults at different rates and consider the different over-provisioning techniques and deployments schemes.
The two node types do not differ in their availability models at the node level because they use the same hardware.
However, the performance achieved by each type is different and we need to use a higher number of $\mathit{ft}$ nodes compared to $\mathit{native}$ nodes to fulfil the same performance constraints.
This results in different availability levels achieved in a cluster by each type of node (see later in \autoref{fig:prism_hw_allNodesUpAvailability}).

\smallskip\noindent\textbf{Passive failover.}
\autoref{fig:prism_availability_stateMachine}(a) presents the state machine diagram for the availability models of PF deployed in the cloud or on-premises.
The PF technique assumes the existence of a pool that has a number of \emph{cold nodes} that are turned off and hence do not fail.
With a cloud deployment, the number of nodes passively waiting in the pool is unlimited ($\mathit{Pool} = \infty$) compared with concrete value ($\mathit{Pool} \geq 0$) for the on-premises case.
The cluster initially consists of $\mathit{UpNodes}$.
We start with a number of nodes that fulfil the performance constraint assuming always available nodes, \ie $\mathit{UpNodes}$ = $\mathit{Num}$ (\raisebox{-1pt}{\ding{202}}).
A fault $\mathit{\lambda}_\mathit{HWCrash}$  crashes a node in the cluster (\raisebox{-1pt}{\ding{203}}), hence reducing the number of $\mathit{UpNodes}$.
Because $\mathit{UpNodes}$ nodes can crash independently, this rate is multiplied by the number of nodes that can be affected by such a fault.
A crashed node can be recovered by replacing it with a node from the pool with a rate of $\mathit{\rho}_\mathit{CrashRecovery}$ (\raisebox{-1pt}{\ding{204}}). 
This rate is affected by the number of available nodes in the pool (\raisebox{-1pt}{\ding{204}} and \raisebox{-1pt}{\ding{205}}).
This would increase the $\mathit{UpNodes}$ while, additionally, decreasing the $\mathit{Pool}$ size in an on-premises deployment.
The crashed nodes from the pool may be repaired (\raisebox{-1pt}{\ding{206}}) and returned back to the pool to use if needed at rate of $\mathit{\rho}_\mathit{PoolRepair}$, which would increase the number of available nodes inside the pool. 
Note that \raisebox{-1pt}{\ding{205}} and \raisebox{-1pt}{\ding{206}} are special cases for the on-premises deployment.

To calculate the availability achieved by such a model, we use ``rewards'' to represent the time spent in each state in a one year time period.
The cluster traverses different states according to the failures and recovery rates, and the time spent in a state where $\mathit{UpNodes}$ = $\mathit{Num}$ represents the cluster availability.
If the time spent outside this state does not exceed the downtime specified by the availability constraints (\eg three 9's availability = 8.77 hours downtime per year), the cluster is considered available and well-performing.

\smallskip\noindent\textbf{Active route anywhere.}
\autoref{fig:prism_availability_stateMachine}(b) presents the state machine diagram for the availability models of ARA deployed in the cloud or on-premises.
In ARA, in addition to the initial number of nodes needed to fulfil performance constraints ($\mathit{Num}$), we additionally use over-provisioned nodes ($\mathit{OP}$) which are actively participating in the cluster.
Therefore, the number of active nodes in the cluster is $\mathit{UpNodes}$ = $\mathit{Num}$ + $\mathit{OP}$ (\raisebox{-1pt}{\ding{202}}).
A fault $\mathit{\lambda}_\mathit{HWCrash}$  crashes a node in the cluster (\raisebox{-1pt}{\ding{203}}), hence reducing the number of $\mathit{UpNodes}$.
Unlike PF, this rate is multiplied by all active nodes including the over-provisioned nodes ($\mathit{UpNodes}$). 
In a cloud deployment (\raisebox{-1pt}{\ding{204}}), the failed node is replaced automatically by another node at rate of $\mathit{\rho}_\mathit{CrashRecovery}$, which increments $\mathit{UpNodes}$.

We calculate the cluster availability by considering the time spent in a state where $\mathit{UpNodes} \geq \mathit{Num}$.
Consequently, at most $\mathit{OP}$ nodes can fail simultaneously without violating the availability and performance constraints.
Note that the cluster can have $\mathit{UpNodes} < \mathit{Num}$ at any time, but it is considered available and well-performing as long as it does not violate the availability constraint.

\begin{figure}[t!]
	\centering
    \includegraphics[scale=0.7]{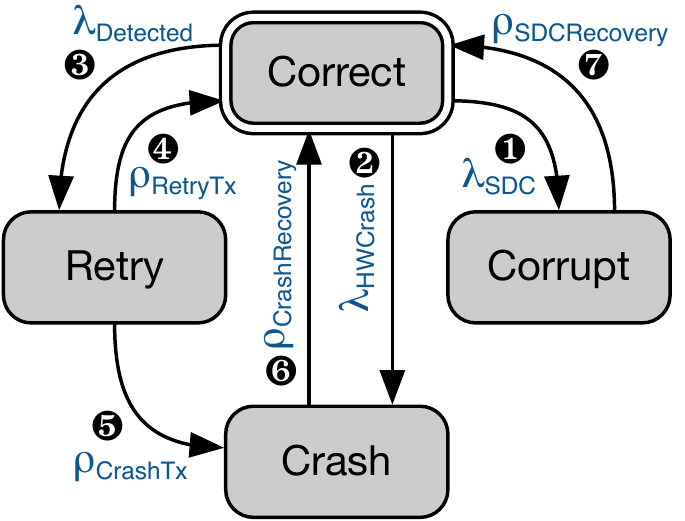}
    \caption{State machine for integrity models of the different nodes.} 
    \label{fig:prism_integrity_stateMachine}
\end{figure}

\smallskip\noindent\textbf{Integrity models.}
In the integrity models, we inject integrity faults considering the different node types and deployments schemes.
$\mathit{ft}$ nodes implementing the HAFT approach have two modes of execution.
HAFT implements instruction level redundancy to detect any violation to computation integrity.
The first mode implements a fail-stop model, specifically, once a violation is detected, computation is stopped ($\mathit{ft}_\mathit{ilr}$).
The second mode targets the availability, specifically, after an integrity failure is detected, instead of aborting, the execution is retried using transactions ($\mathit{ft}_\mathit{tx}$).
\autoref{fig:prism_integrity_stateMachine} presents the state machine diagram for integrity models of a node deployed in the cloud or on-premises.
The node starts with $\mathit{Correct}$ state.
A transient fault can result in corruption of the state (SDC), crash of the application or masking of the fault.
If not masked, a transient fault transfers the node in a $\mathit{Corrupt}$ state at rate of $\mathit{\lambda}_\mathit{SDC}$ (\raisebox{-1pt}{\ding{202}}) or in a $\mathit{Crash}$ state at rate of $\mathit{\lambda}_\mathit{HWcrash}$ (\raisebox{-1pt}{\ding{203}}).
Note that using $\mathit{ft}_\mathit{ilr}$ nodes, this rate also includes the crashes resulting from aborting the application after detecting a transient fault. 
A $\mathit{ft}_\mathit{tx}$ node detects transient faults at rate of $\mathit{\lambda}_\mathit{Detected}$ and transfers the node into $\mathit{Retry}$ state (\raisebox{-1pt}{\ding{204}}). 
A node in a $\mathit{Retry}$ state is able to either recover the state at rate of $\mathit{\rho}_\mathit{RetryTx}$ and revert the node back into $\mathit{Correct}$ state (\raisebox{-1pt}{\ding{205}}), or if retry is not successful, abort execution and transfer the node into $\mathit{Crash}$ state at rate of $\mathit{\rho}_\mathit{CrashTx}$  (\raisebox{-1pt}{\ding{206}}).
Both types of nodes do not have any mechanism to recover from crashed states.
However, if deployed in the cloud, a node with $\mathit{Crash}$ state is automatically replaced by another node to match the SLA agreement at a rate of $\mathit{\rho}_\mathit{CrashRecovery}$ (\raisebox{-1pt}{\ding{207}}).
When deployed on-premises, assuming enough resources in the pool, the node transfers back to $\mathit{Correct}$ state.
Additionally, the corruption of a node may be manually detected at a rate of $\mathit{\rho}_\mathit{SDCRecovery}$, which reverts the node back to $\mathit{Correct}$ state (\raisebox{-1pt}{\ding{208}}). 

The integrity level achieved is captured by measuring the normalised time spent by the node in each state: $\mathit{Correct}$ as correct time, $\mathit{Corrupt}$ as corrupted time, $\mathit{Crash}$ combined with $\mathit{Retry}$ as downtime.
Note that over-provisioning of nodes does not help to reduce the time in $\mathit{Corrupt}$ state, since it produces multiple nodes with similar integrity behaviour.
Alternatively, a different integrity protection technique should be implemented at the node level.


\section{Evaluation}
\label{sec:evaluation}

In this section we illustrate how \sys's two phase methodology can be used.
In a first step we benchmark a single server node deployed with $\mathit{native}$, $\mathit{ft}_\mathit{ilr}$ and $\mathit{ft}_\mathit{tx}$ nodes using a stateless web server application and two soft-state applications.
Then, we parameterise the models built in \sys to calculate the availability and the integrity behaviour at the cluster level and decide the required capacity.

\begin{figure*}[t!]
    \centering
    \includegraphics[scale=0.7]{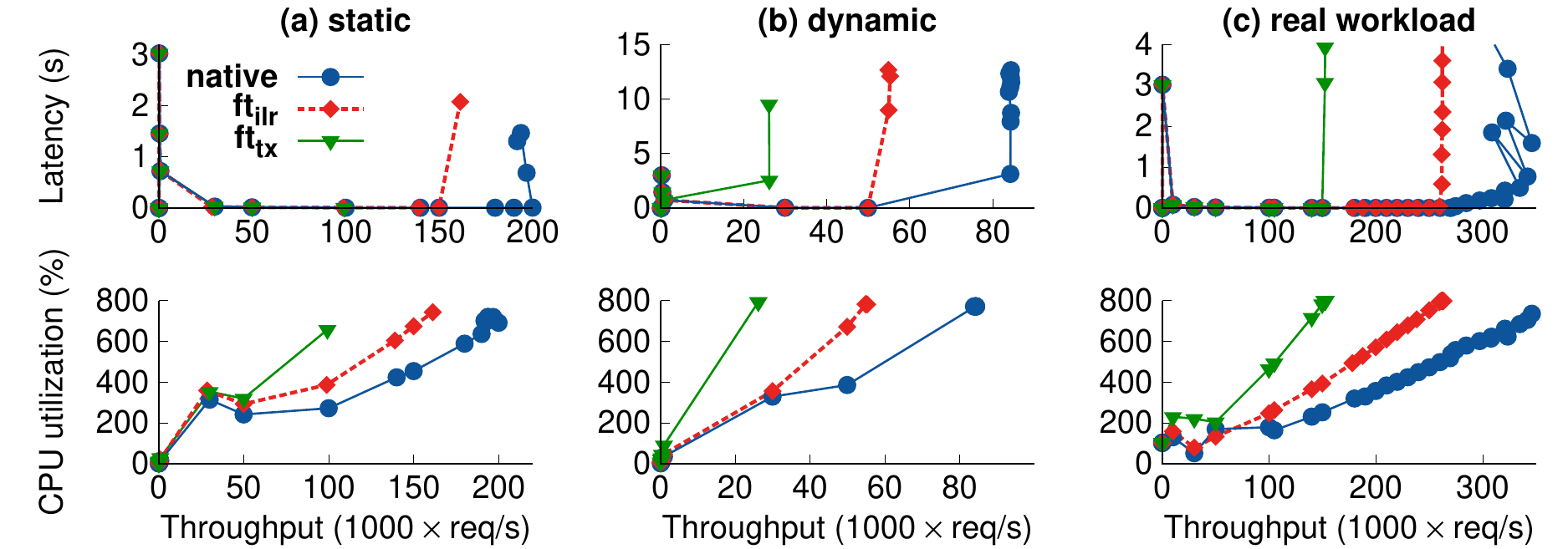}
    \caption{Apache throughput vs. latency (top) and CPU utilisation (bottom) with 3 different workloads: static (a), dynamic (b) and real-world (c).} 
    \label{fig:latency-throughput-apache}
\end{figure*}

\begin{figure}[t!]
    \centering
    \includegraphics[scale=0.7]{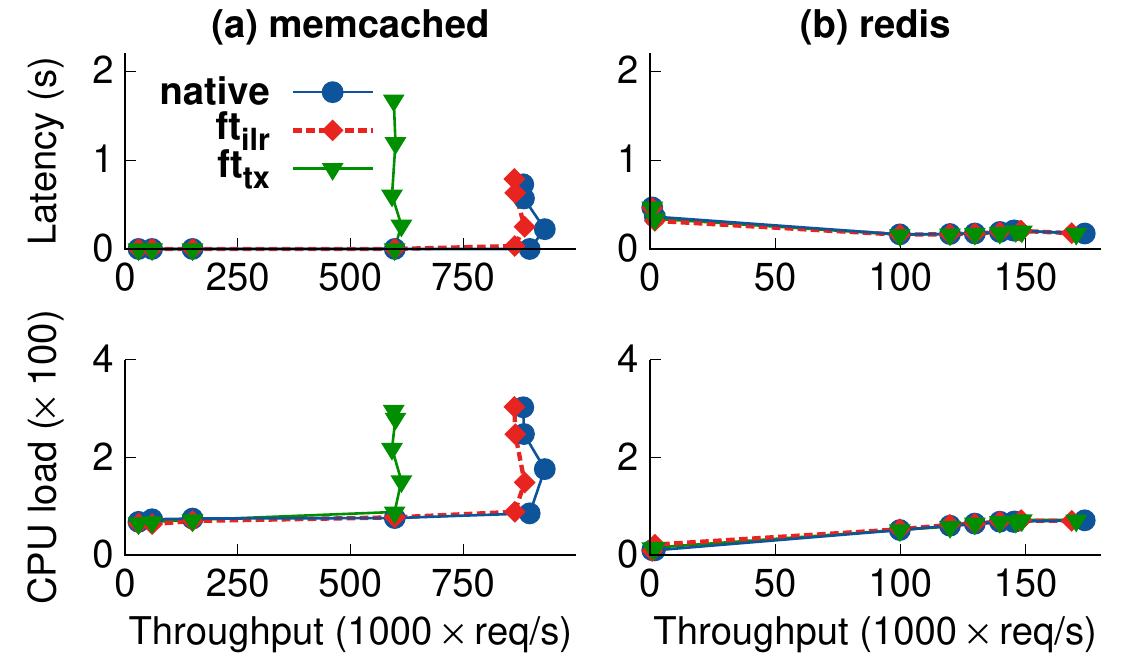}
    \caption{Memcached~(a) and Redis~(b): throughput vs. latency (top) and CPU load (bottom).}  
    \label{fig:latency-throughput-memcached}
\end{figure}

Our evaluation answers the following questions: 
\begin{enumerate*}[label=(\arabic*)]
    \item How much performance, \ie throughput and latency, can be achieved by a single node of types $\mathit{native}$, $\mathit{ft}_\mathit{ilr}$ and $\mathit{ft}_\mathit{tx}$?
    \item What is the effect of the deployment scheme (cloud vs. on-premises) on the availability level achieved by a single node, without over-provisioning?
    \item What is the effect of using a cluster of nodes on the availability achieved, without over-provisioning?
    \item How many extra nodes are needed to ensure a given availability and performance constraints when using ARA and PF with different deployment schemes?
    \item How a single node of types $\mathit{native}$, $\mathit{ft}_\mathit{ilr}$ and $\mathit{ft}_\mathit{tx}$ deployed in the cloud would behave when transient faults exist?
\end{enumerate*}

\subsection{Experimental settings}

All experiments use a dedicated on-premises deployment.
The experimental part attempts to identify the performance of a single node $\mathit{NodT}$, which can then be used to identify the number of nodes ($\mathit{Num}$) required to achieve the cluster target performance ($\mathit{SerT}$) assuming always the available nodes.
As an actual cloud deployment might differ experimentally in the performance achieved by each node $\mathit{NodT}$, it can use a similar experimental path as presented here to obtain $\mathit{Num}$.
In this paper, we want to study the pure effect of over-provisioning and node types on the availability of the cluster.
We therefore assume $\mathit{NodT}$ to be the same for nodes deployed in the cloud or on-premises, and we use the same number of nodes ($\mathit{Num}$) in each deployment to achieve $\mathit{SerT}$.

Each server node has an Intel Xeon E3-1270\,v5 CPU clocked at 3.6\,GHz.
The CPU has 4 cores with 8 hyper-threads (2 per core) and 8\,MB of L3-cache.
The server has 64\,GB memory running Ubuntu 14.04.5 LTS with Linux 4.4.
The workload generators run on a server with two 14-core Intel Xeon E5-2683\,v3 CPUs at 2\,GHz with 112\,GB of RAM and Ubuntu 15.10.
All machines have a 10\,Gb/s Ethernet NIC connected to a dedicated network switch.

We use the following three real-world applications: Apache web server~(v2.2.11) \cite{httpd}, memcached~(v1.4.21) \cite{memcached} and Redis~(v2.8.7) \cite{redis}.
Furthermore, we deploy the applications in three variants: $\mathit{native}$ unmodified application, $\mathit{ft}_\mathit{ilr}$ built using the HAFT LLVM tool chain \cite{Kuvaiskii2016HAFT} with ILR, and $\mathit{ft}_\mathit{tx}$ which additionally execute the application inside transactions.

\subsection{Experimental measurements}

\smallskip\noindent\textbf{Apache Web Server.}
Our first set of experiments study the Apache \texttt{httpd} web server (\autoref{fig:latency-throughput-apache}).
We evaluate the throughput vs. latency ratio and the overall CPU usage.
We first measure the achievable throughput and latency of the three node variants under test, and the CPU utilisation during the execution of the benchmarks.
We use the \emph{wrk2} workload generator \cite{wrk2} to measure the throughput and latency based on fixed request rates issued to the master server.
The following three workloads are used.
\begin{enumerate*}[label=\emph{(\roman*)}]
    \item \emph{static content:} the web server fetches static content (such as images or CSS files);
    \item \emph{dynamic content:} the web server fetches dynamic content generated by PHP scripts~(v5.4.0); and
    \item \emph{real workload:} the web server operates under real-world conditions by retrieving a WordPress blog page with a MySQL server database at the back-end.
\end{enumerate*}
We gradually increase the submission rate of HTTP requests until the response times hit unacceptable levels (\eg $>1$ second).

\autoref{fig:latency-throughput-apache}a shows the results for static content.
The x-axis shows the measured response rate of the submitted requests while the y-axis shows the corresponding latency.
The measured response rate is often lower than the introduced request rate when the system is saturated, giving the impression of the line going backwards suddenly since the data points are sorted by the introduced request rate which is constantly increasing, as seen in top graphs of \autoref{fig:latency-throughput-apache}.

For dynamic content, \autoref{fig:latency-throughput-apache}b depicts the results to fetch a PHP script that solely employs an empty for-loop iterating $10^3$ times to basically simulate some CPU-intensive workload.
Lastly, \autoref{fig:latency-throughput-apache}c presents the results under real workload for blog-like web pages.
As expected, the $\mathit{native}$ execution outperforms the other variants, with 200 thousands requests per seconds (kreq/s) to fetch static content.
With $\mathit{ft}_\mathit{tx}$ we reach only half the throughput, \ie 100\,kreq/s, whereas $\mathit{ft}_\mathit{ilr}$ improves the performance (up to 155\,kreq/s for static content) but without the ability to properly handle detected errors.
For dynamic content, we observe similar trends.
The $\mathit{native}$ mode excels with a peak of roughly 86\,kreq/s, followed by $\mathit{ft}_\mathit{ilr}$ and $\mathit{ft}_\mathit{tx}$, respectively at 52 and 23\,kreq/s. 
Surprisingly, for blog-like web pages, the peak performance for both $\mathit{native}$ and $\mathit{ft}$ is much higher than when solely retrieving static content or a plain PHP page without any database interaction.
We observe a peak performance of 321\,kreq/s for $\mathit{native}$, and respectively 260 and 160\,kreq/s for the two $\mathit{ft}$ variants.
This effect can be explained by lower thread contention upon database lookups happening for each request.

In terms of CPU utilisation, as expected we observe an increase as more requests are being issued.
In general, all the variants are strictly CPU-bound (the limiting factor is our hardware) and the injected workloads manage to fully saturate all CPU cores.
However, the $\mathit{ft}$ variants saturate the CPU much more quickly than the $\mathit{native}$ execution.
For example, with static content we reach roughly the 800\% CPU limit with a throughput of around 200\,req/s.
This is confirmed by a corresponding increase in response latency.
This behaviour is expected since $\mathit{ft}$ requires more CPU cycles (for the instrumented instructions) and thus saturates the CPU more quickly.
Similarly, CPU consumption also increases for the benchmarks that retrieve dynamic content, either with and without database interaction.

\smallskip\noindent\textbf{Key-value stores: memcached and Redis.}
Next, we evaluate two widely-used key-value stores: memcached and Redis.
To measure throughput vs. latency with memcached, we rely on Twitter's \emph{mcperf} \cite{mcperf} tool.
For Redis, we use YCSB \cite{Cooper:2010:BCS:1807128.1807152} with workload A, which comprises $50\%$ read and $50\%$ update operations.
\autoref{fig:latency-throughput-memcached} shows the results for both systems.

Memcached reaches a peak throughput at 886\,kreq/s for $\mathit{native}$ and 600\,kreq/s for $\mathit{ft}_\mathit{tx}$. 
The $\mathit{ft}_\mathit{ilr}$ variant is close to $\mathit{native}$, with only a 12\% difference overall.
Since memcached is limited by the memory bandwidth (8\,GB/s on our hardware), there is only a small increase in CPU utilisation once the system is saturated.

With Redis, we observe a peak at around 120\,kreq/s before the latency starts climbing.
Interestingly, there is almost no visible overhead for the $\mathit{ft}$ variants in comparison to $\mathit{native}$.
This is because Redis is single-threaded while $\mathit{ft}$ harnesses multi-core technology for ILR, as confirmed by the fact that CPU usage slightly increases as more requests are being processed, yet never exceeds 100\% (\ie 1 core).

In summary, we observe across all applications an average throughput ($\mathit{NodT}$) of about 71\% for $\mathit{ft}_\mathit{tx}$ in comparison to $\mathit{native}$ execution.
If we use $\mathit{ft}_\mathit{ilr}$, the throughput climbs up to 92\% of $\mathit{native}$ performance.

\begin{table*}[t!]
\centering
\begin{subtable}[t]{.5\linewidth}
    \small
    \centering
	\begin{tabular}{l|c|c|c}
	\textbf{Transient faults} & $\mathit{native}$ & $\mathit{ft}_\mathit{ilr}$ & $\mathit{ft}_\mathit{tx}$ \\
	\hline
	Corrupt (\%)     &  26.19   &  0.8    &  1.17    \\                   
	Crash (\%)       &  12.49   &  75.0   &  7.72     \\                  
	Retry (\%)       &  --      &  --     &  66.99    \\
	\hline
	\end{tabular}
\end{subtable}%
\begin{subtable}[t]{.5\linewidth}
    \small
    \centering
	\begin{tabular}{l|c|c|c}
	\textbf{Recovery time} & $\mathit{native}$  & $\mathit{ft}_\mathit{ilr}$  & $\mathit{ft}_\mathit{tx}$ \\
	\hline
	\rule{0pt}{1em}Crash recovery (s)   &  \multicolumn{3}{c}{15\textsuperscript{\raisebox{-1pt}{a}}} \\
	SDC recovery (h)  & \multicolumn{3}{c}{ 6\textsuperscript{\raisebox{-1pt}{b}}} \\
    \cline{2-4}
	Retry transaction ($\mu$s)         & --         & --       &   2.5\textsuperscript{\raisebox{-1pt}{c}} \\
	\hline
	\end{tabular}
\end{subtable}
\smallskip
\caption{Probabilistic models parameters: transient fault probabilities \cite{Kuvaiskii2016HAFT} (left) and recovery times (right).\\[3pt]
\scriptsize
\textsuperscript{(a)}Common values for failover in HA cluster is 1-30\,s \cite{Failover}.
\textsuperscript{(b)}Amazon reported 6\,h to manually recover from corrupted state \cite{AmazonS32008}.
\textsuperscript{(c)}Maximum latency of transaction retry with 5,000 instructions (2.0\,GHz CPU). \label{tab:prism_parameters}}
\end{table*}

\subsection{Dependability evaluation}

We next use the models described in \S\ref{sec:implementation} to evaluate the availability and integrity under faults for single nodes and a cluster of nodes, to calculate the required capacity.

\smallskip\noindent\textbf{Models settings.}
\autoref{tab:prism_parameters} presents the main parameters used in the models.
The table consists of two parts.
The first part presents the probabilities of state transitions in the integrity models from $\mathit{Correct}$ state to any of the other states ($\mathit{Corrrupt}$, $\mathit{Crash}$, $\mathit{Retry}$) when transient faults are injected in a node of a given type ($\mathit{native}$, $\mathit{ft}_\mathit{ilr}$ and $\mathit{ft}_\mathit{tx}$). 
These values are produced using fault injection experiments on a wide range of applications in the HAFT paper \cite{Kuvaiskii2016HAFT} (Table 4). 
The second part presents the time needed to recover the different failed states back to $\mathit{Correct}$ state.
The techniques include replacing a crashed node by either failover to a new node ($\mathit{CrashRecovery}$), the manual repair of a node with SDC state ($\mathit{SDCRecovery}$) and the recovery of a node with a detected transient fault by retrying a transaction ($\mathit{RetryTx}$).
Note that recovery times can be converted into rates assuming the second (s) as the basic time unit, using $\mathit{RecoveryRate = 1/RecoveryTime}$.
In the rest of this section we use the values presented in the table unless otherwise specified.

\begin{figure}[t!]
    \centering
    \includegraphics[scale=0.7]{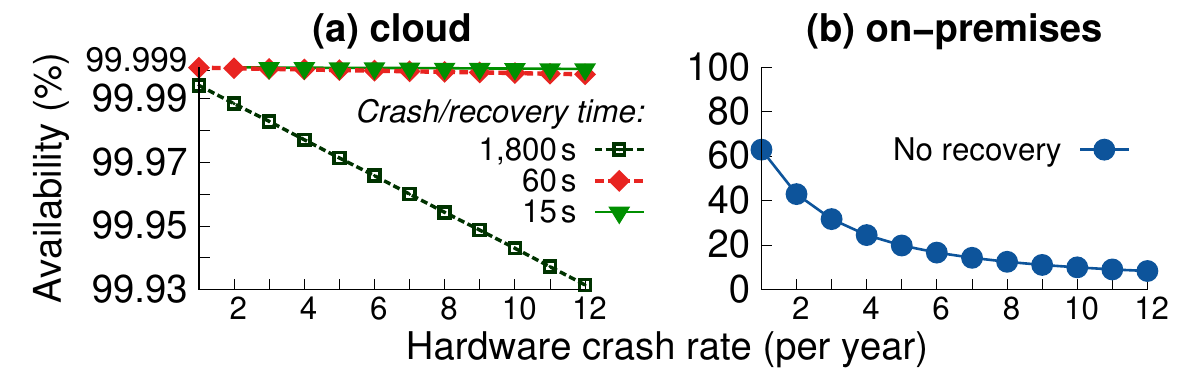}
    \caption{Availability of a single-node for one year when deployed in the cloud (a) and on-premises (b), without over-provisioning.}
    \label{fig:single_node_deployment}
\end{figure}

\smallskip\noindent\textbf{Availability of single node deployment.}
We want to study the availability achieved by deploying a single node without any over-provisioning techniques in the cloud compared to on-premises.
The cloud differs from on-premises deployment in its ability to automatically fail over by replacing a failed node with another one from a hypothetically unlimited pool, so as to satisfy the SLA agreement.
The cloud automatic failover is modelled with $\mathit{Pool} = \infty$ and three values for $\mathit{CrashRecovery} = $\{15\,s, 60\,s, 1800\,s\}, while on-premises uses $\mathit{Pool} = 0$.
Both deployments use $\mathit{Num} = 1$ to model a single node in the cluster. 
\autoref{fig:single_node_deployment} shows the results of injecting hardware crash faults with rates from once to 12 per year (x-axis) and the availability achieved in one year (y-axis) for a node deployed in cloud (a) and on-premises (b).
The availability calculated represents the operational availability $\mathit{upTime/totalTime}$, where \emph{up} time assumes that $Num$ nodes are operational and \emph{total} time is one year.

\begin{figure}[t!]
    \centering
    \includegraphics[scale=0.7]{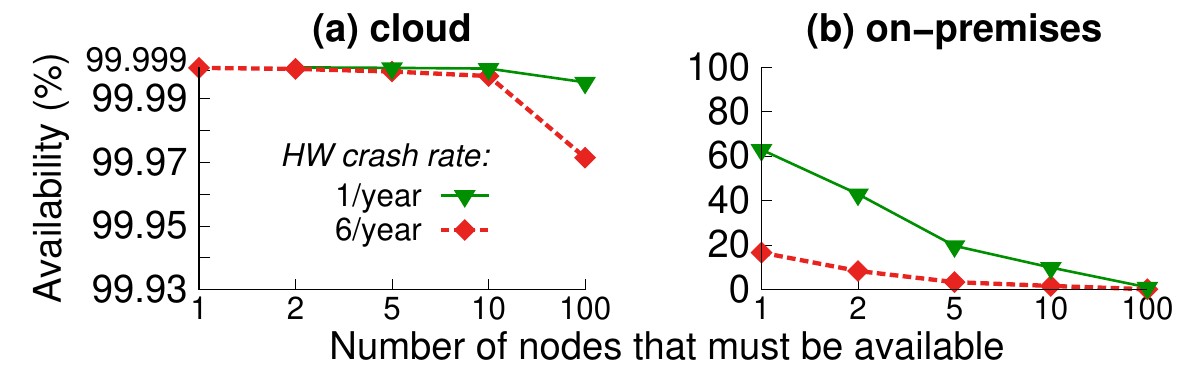}
    \caption{Availability of a service under a given crash fault rates and without over-provisioning.}
    \label{fig:prism_hw_allNodesUpAvailability}
\end{figure}

\autoref{fig:single_node_deployment}\,(a) shows that the cloud deployment can achieve at least five 9's for a single node in terms of yearly availability.
With slower recovery time (1800\,s), the achieved availability of at least three 9's is consistent with what most current cloud providers would provide for a single node \cite{AmazonSLA3Nines}.
\autoref{fig:single_node_deployment}\,(b) shows that on-premises cannot achieve the required availability levels even with very low fault rate, and availability quickly degrades with higher fault rates.
The faster degradation in availability is due to missing repairs in the on-premises deployment.

\begin{table*}[t!]
\centering
\begin{subtable}[b]{.55\linewidth}
    \small
    \centering
	\begin{tabular}[b]{l|c|rc|c|c|c|c|c}
        {} & {} & {} & \multicolumn{6}{c}{\textbf{Extra ARA nodes}} \\
        {} & {} & \textbf{HCR:} & \multicolumn{3}{c|}{\textbf{1/year}}& \multicolumn{3}{c}{\textbf{6/year}} \\
        {\textbf{Node}} & \textbf{Base} & \textbf{CR:} & \textbf{15\,s} & \textbf{1\,min} & \textbf{30\,min} & \textbf{15\,s} & \textbf{1\,min} & \textbf{30\,min} \\
        \hline
        $\mathit{native}$          & 10 & & 0 & 0 & 0 & 0 & 0 & 1\\
        $\mathit{ft}_\mathit{ilr}$ & 11 & & 0 & 0 & 0 & 0 & 0 & 1\\
        $\mathit{ft}_\mathit{tx}$  & 15 & & 0 & 0 & 0 & 0 & 0 & 1\\
        \hline
	\end{tabular}
    \subcaption{cloud}
\end{subtable}%
\hfill
\begin{subtable}[b]{.35\linewidth}
    \small
    \centering
	\begin{tabular}[b]{l|c|rc|c}
        {} & {} & \multicolumn{3}{r}{\textbf{Extra ARA nodes}} \\
        \textbf{Node} & \textbf{Base} & \textbf{HCR:} & \textbf{1/year} & \textbf{6/year} \\
        \hline
        $\mathit{native}$          & 10 & & 35 & 113\\
        $\mathit{ft}_\mathit{ilr}$ & 11 & & 37 & 121\\
        $\mathit{ft}_\mathit{tx}$  & 15 & & 46 & 152\\
        \hline
	\end{tabular}
    \subcaption{on-premises}
\end{subtable}
\caption{Number of active nodes needed (base + extra) for \emph{cloud} and \emph{on-premises} deployments with ARA to achieve 10$\times$ native throughput ($\mathit{SerT}$) and three 9's availability level (HCR=$\mathit{\lambda}_\mathit{HWcrash}$, CR=$\mathit{\rho}_\mathit{CrashRecovery}$). \label{tab:cloud-self-r10}}
\end{table*}

\begin{table*}[t!]
\centering
\small
    \begin{tabular}{l|c|rc|c|c|c|c|c|c|c|c|c|c|c}
        {} & \textbf{\multirow{4}{*}{Base}}  & {} &\multicolumn{12}{c}{\textbf{Extra pool nodes}} \\
        {} & {} & \textbf{HCR:} & \multicolumn{6}{c|}{\textbf{1/year}} & \multicolumn{6}{c}{\textbf{6/year}} \\
        {\textbf{Node}} & {} & \textbf{CR:} & \multicolumn{2}{c|}{\textbf{15\,s}} & \multicolumn{2}{c|}{\textbf{1\,min}} & \multicolumn{2}{c|}{\textbf{30\,min}} & \multicolumn{2}{c|}{\textbf{15\,s}} & \multicolumn{2}{c|}{\textbf{1\,min}} & \multicolumn{2}{c}{\textbf{30\,min}} \\
        {} & {} & \textbf{PR:} & \textbf{--} & \textbf{1\,h} & \textbf{--} & \textbf{1\,h} & \textbf{--} & \textbf{1\,h} & \textbf{--} & \textbf{1\,h} & \textbf{--} & \textbf{1\,h} & \textbf{--} & \textbf{1\,h} \\
        \hline
        $\mathit{native}$          & 10 & & 18  & 1 & 18 & 1 & 19 & 1 & 30 & 1 & 30 & 1 & $\times$  & $\times$ \\
        $\mathit{ft}_\mathit{ilr}$ & 11 & & 19 & 1 & 19 & 1 & 20  & 1 & 33 & 1 & 33 & 1 & $\times$ & $\times$ \\
        $\mathit{ft}_\mathit{tx}$  & 15 & & 24 & 1 & 24 & 1 & 27 & 1 & 42 & 1 & 42  & 1 & $\times$ & $\times$ \\
        \hline
	\end{tabular}
\smallskip
\caption{Number of nodes needed (base + pool) for \emph{on-premises} deployment with PF to achieve 10$\times$ native throughput ($\mathit{SerT}$) and three 9's availability (HCR=$\mathit{\lambda}_\mathit{HWcrash}$, CR=$\mathit{\rho}_\mathit{CrashRecovery}$, PR=$\mathit{\rho}_\mathit{PoolRepair}$). \label{tab:PF_r10}}
\end{table*}

\smallskip\noindent\textbf{Service target throughput with failure free nodes.}
In the experimental phase, we measured throughput of a single node ($\mathit{NodT}$) for the different node types and calculated the average throughput degradation for $\mathit{ft}_\mathit{ilr}$ and $\mathit{ft}_\mathit{tx}$ with respect to a $\mathit{native}$ node.
By defining the $\mathit{SerT}$ as multiples of $\mathit{NodT}_\mathit{native}$, we can calculate the number of nodes $\mathit{Num}$ needed to achieve a service target throughput for all node types using  $\mathit{SerT}/\mathit{NodT}$.
For example, for $\mathit{SerT}=1$, we need either one $\mathit{native}$ node or two nodes of either $\mathit{ft}_\mathit{ilr}$ or $\mathit{ft}_\mathit{tx}$ type.
Similarly, we need either 10 $\mathit{native}$, 11 $\mathit{ft}_\mathit{ilr}$ or 15 $\mathit{ft}_\mathit{tx}$ nodes to handle $\mathit{SerT}=10$.
The given nodes are considered sufficient to fulfil the performance constraints of the service under the assumptions that
\begin{enumerate*}[label=\emph{(\roman*)}]
    \item nodes do not fail and
    \item the backend infrastructure scales with the number of nodes in the cluster.
\end{enumerate*}
If the backend infrastructure does not scale with the number of nodes, contention will increase on the backend resources, hence reducing $\mathit{NodT}$ and increasing the response time.
In this case, more nodes are required to achieve the same $\mathit{SerT}$, which should then be determined experimentally.

\smallskip\noindent\textbf{Availability of multiple node deployments.}
The service requires $Num$ of nodes to achieve the performance constraints assuming always available nodes.
Nodes do, however, crash in practice, which leads to degraded service performance.
We want to study the effect of nodes crashing on the overall service availability using different deployments schemes when no over-provisioning techniques are used.
In \autoref{fig:prism_hw_allNodesUpAvailability}, we vary the number of nodes in the cluster (x-axis) when deployed in the cloud (a) and on-premises (b), and calculate the availability achieved at the cluster level (y-axis) when hardware crash faults are injected with two rates (1 and 6 per year).
We measure the availability of cluster with $UpNodes = Num$ during one year, \ie the yearly percentage of time that the service is able to fulfill its performance requirement.
We observe that, as we increase the number of nodes expected to be operational, the availability of the service deployed in the cloud remains high, while it degrades fast on-premises even with low fault rate.
Therefore, deploying a service on-premises requires over-provisioning to ensure that performance requirement are met.

\smallskip\noindent\textbf{Capacity for dependable service.}
A dependable service that relies on a cluster of nodes to achieve a target QoS requires the cluster to satisfy availability constraints, in addition to protecting the integrity of the service.
To that end, we can use over-provisioning (\eg PF or ARA) to meet the availability objectives as well as integrity protection techniques to enhance the integrity of the service (\eg HAFT).
We use the capacity planning process in two ways: 
\begin{enumerate*}[label=\emph{\roman*)}]
    \item to define the number of nodes needed for a dependable service built using different node types, deployment schemes and over-provisioning techniques, and
    \item to quantify the integrity of the service.
\end{enumerate*}

\smallskip\textbf{(1) Number of nodes needed.}
For \emph{cloud deployments}, \autoref{tab:cloud-self-r10}\,(a) shows the number of nodes needed for a dependable service that has a performance constraint $\mathit{SerT} = 10 \times \mathit{NodT}_\mathit{native}$ and service availability of at least three 9's.
We consider $\mathit{CrashRecovery}$ times for the automatic failover as 15 seconds, 1 minute and 30 minutes, and hardware crash rates of 1 and 6 per year per node.
If the availability achieved does not satisfy the required level, extra nodes are provisioned as ARA nodes.
The deployment types differ in their ``base'' number of nodes, yet they all achieve high availability levels of at least three 9's except with high crash rate and slow failover time (HCR = 6/year and CR = 30\,min).
In this case, it is enough to have a single extra ARA node over-provisioned in the cloud to meet the required availability level.

For \emph{on-premises deployment}, we consider both ARA in \autoref{tab:cloud-self-r10}\,(b) and PF in \autoref{tab:PF_r10} as over-provisioning techniques.
The tables consider a service required throughout $\mathit{SerT} = 10 \times \mathit{NodT}_\mathit{native}$ and a service availability of at least three 9's.
\autoref{tab:cloud-self-r10}\,(b) shows that the dependable service requires a number of active nodes $\mathit{UpNodes}$ equal to the sum of the ``base'' nodes needed for performance and the over-provisioned ARA nodes.
For example, a service that requires 10 $\mathit{native}$ nodes also needs, assuming a crash rate of 1 per year, 35 additional active nodes to ensure three 9's availability.
To understand this high number of over-provisioned nodes, consider that an on-premises deployment can achieve only 10\% availability with a cluster of 10 nodes (\autoref{fig:prism_hw_allNodesUpAvailability}), which is very low compared to the required level of 99.9\%.
Additionally, ARA nodes are active nodes and can fail due to crash failures.
Therefore, we need to over-provision many nodes to ensure that at least 10 are available at any time, except for the allowed downtime (8.77 hours per year for three 9's).
Note that the number of base nodes for each type is different, which also affects the number of additional ARA nodes.

\autoref{tab:PF_r10} shows that a dependable service requires a number of nodes equal to the sum of ``base'' active node needed for performance and the over-provisioned nodes as PF.
The table considers $\mathit{CrashRecovery}$ times for the failover from the pool as 15 seconds, 1 minute and 30 minutes and hardware crash rates of 1 and 6 per year per node.
Additionally, we consider two cases for handling crashed nodes upon failed-over to a node from the pool: ``no repair'' (denoted by \textbf{--}) and ``1 hour repair''.
If repaired, a crashed node can return to the pool and be used for further failover upon need.
The table indicates that the number of nodes required to fulfil the availability constraints is dramatically reduced when repairing crashed nodes, as compared to the ``no repair'' case.
For example, for a crash rate of 1 per year and 15\,s failover time, a service would require, in order to achieve three 9's availability using 10 $\mathit{native}$ active nodes, 18 additional passive nodes in the pool if there are no repair vs. one if the pool is repaired at a rate of one per hour.

\begin{figure}[t!]
\centering
\includegraphics[scale=0.7]{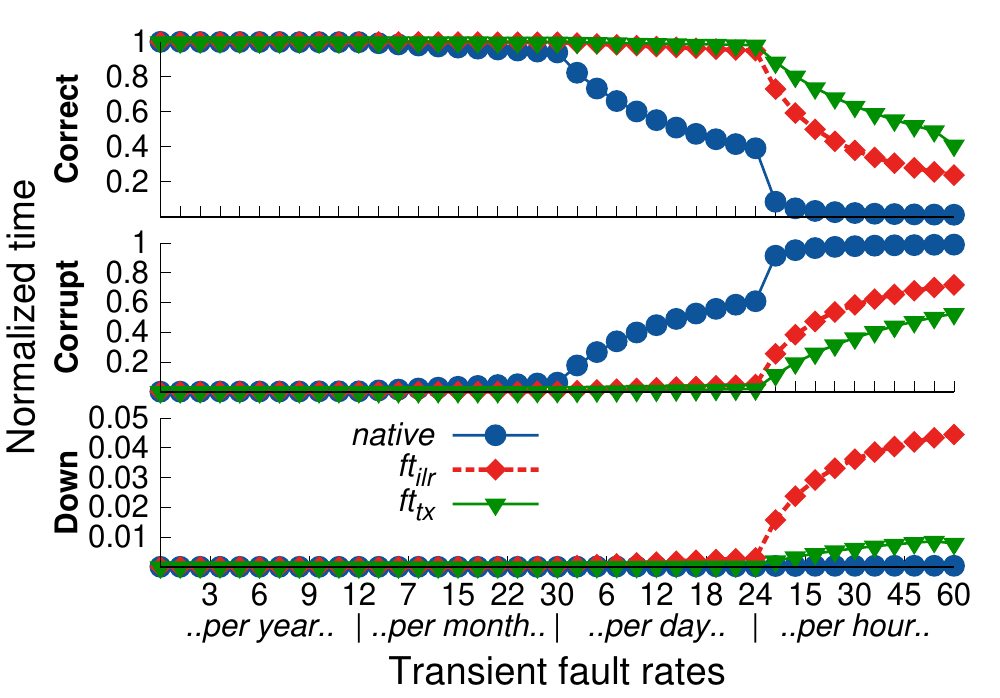}
\caption{Normalised time spent by the different node types in correct, corrupted and down states after injecting transient faults.}  
\label{fig:prism_tf_UpSdcDownTime}
\end{figure}

Comparing \autoref{tab:cloud-self-r10}\,(b) and \autoref{tab:PF_r10}, one can see that PF requires fewer nodes than ARA.
Indeed, passive nodes in the pool work in cold mode (turned off) and not exposed to crashes.

\textbf{(2) Integrity of the service.}
To study how different nodes behave when transient faults are injected, we use the integrity models presented in \autoref{fig:prism_integrity_stateMachine}.
If not masked, a transient fault can crash the node or corrupt its internal state with probabilities that vary between different node types due to their ability to tolerate transient faults, as seen in \autoref{tab:prism_parameters}.
Note that $\mathit{native}$ nodes do not implement any integrity protection mechanism, while $\mathit{ft}$ nodes are hardened against data corruption.
Transition rates between $Correct$ state and other states ($\mathit{Corrupt}$, $\mathit{Crash}$, $\mathit{Retry}$) are defined by the injected transient faults rate multiplied by the corresponding probability.

\autoref{fig:prism_tf_UpSdcDownTime} presents the normalised time that each node type spends in $\mathit{Correct}$ (available), $\mathit{Corrupt}$ (available but integrity is not preserved) and $\mathit{Down}$ (unavailable, crashed or under repair) states in one month in the cloud.
The figure shows that with low transient fault rates, all nodes spend most of their time in $\mathit{Correct}$ state.
By increasing fault rates, nodes spends more time in $\mathit{Corrupt}$ or $\mathit{Crash}$ states.
Specifically, $\mathit{native}$ nodes spend 0.2--5.5\% of time in $\mathit{Corrupt}$ state with faults in the month range and 6--58\% with faults in the day range, whereas these percentages decrease to 0.007--0.18\% and 0.2--4.3\% with $\mathit{ft}_\mathit{ilr}$ and 0.0026--0.07\% and 0.07--1.67\% with $\mathit{ft}_\mathit{tx}$ for the same ranges, respectively.
With high fault rates, $\mathit{ft}_\mathit{tx}$ spends more time than $\mathit{ft}_\mathit{ilr}$ in $\mathit{Correct}$ state and less in $\mathit{Corrupt}$ or $\mathit{Down}$ states.
This is due to the use of transactions to recover from a detected fault in $\mathit{ft}_\mathit{tx}$, as compared to the slower failover recovery in $\mathit{ft}_\mathit{ilr}$.
Therefore, $\mathit{ft}_\mathit{tx}$ is not only more reliable, but also more available than the other node types.
Service dependability should not only consider the availability of the nodes in the cluster, but also the integrity of the available nodes, since they may spend a considerable amount of time in $\mathit{Corrupt}$ state if integrity mechanisms are not implemented.


\section{Conclusion} 
\label{sec:conclusion}

We have developed a capacity planning process (\sys) to quantify the number of nodes needed to ensure the dependability of stateless services.
We consider availability, integrity and performance failures for cloud-based and on-premises deployments.
\sys combines a two-phase process---empirical and modelling-based.
In the \emph{empirical phase}, we characterise the performance and integrity of the service at the node level to parameterise the \emph{modelling phase}, in which we implement probabilistic models to estimate the availability and integrity of service.
Our evaluation of \sys using Apache, memcached and Redis shows that both availability and performance are important to leverage the benefits of dependability mechanisms.

\newpage

\balance

\bibliographystyle{IEEEtran}
\bibliography{compact}

\end{document}